\documentclass[aps,prl,twocolumn,groupedaddress,showpacs,floatfix,superscriptaddress,longbibliography]{revtex4-2}
\usepackage[plainpages=false,pdfpagelabels,colorlinks=true,linkcolor=red,urlcolor=blue,citecolor=blue,pdftitle={Title},pdfauthor={},pdfdisplaydoctitle=true,pdfduplex=DuplexFlipLongEdge]{hyperref}
\usepackage{epsfig}
\usepackage{graphicx}
\usepackage[utf8]{inputenc}
\usepackage{amsmath,amssymb}
\usepackage[dvipsnames]{xcolor}

\usepackage[normalem]{ulem}

\begin{document}

\title{Flat bands and superconductivity induced by periodic strain in monolayer graphene}
\author{Jingyao Meng}
\affiliation{School of Physics and Astronomy, Beijing Normal University, Beijing 100875, China\\}
\author{Runyu Ma}
\affiliation{School of Physics and Astronomy, Beijing Normal University, Beijing 100875, China\\}
\author{Tianxing Ma}
\email{txma@bnu.edu.cn}
\affiliation{School of Physics and Astronomy, Beijing Normal University, Beijing 100875, China\\}
\affiliation{Key Laboratory of Multiscale Spin Physics (Ministry of Education), Beijing Normal University, Beijing 100875, China\\}
\author{Hai-Qing Lin}
\affiliation{School of Physics and Institute for Advance Study in Physics, Zhejiang University, Hangzhou 310058, Zhejiang, China}
\affiliation{Beijing Computational Science Research Center, Beijing 100193, China}
\affiliation{School of Physics and Astronomy, Beijing Normal University, Beijing 100875, China\\}

\begin{abstract}
Superconductivity in single-layer graphene has attracted considerable interest. Here, using the determinant quantum Monte Carlo method, we study transitions of superconductivity and magnetism in a monolayer graphene with a special periodic strain.
Consistent with experiments,
the deformation accumulates a series of flat bands, whose robustness under interaction is verified through electron localization in real space.
During the reconstruction of the band structure,
the superconductivity appears in flat band range with next-nearest neighbor $d+id$ pairing symmetry dominating other modes
and is accompanied by ferromagnetism caused by symmetry breaking.
We also demonstrate that the strain-induced symmetry breaking would accumulate an energy-gap antiferromagnetic insulating phase at half filling even under the limitation of not strong enough interaction, which shows its potential as a platform that exhibits strongly correlated phenomena.
\end{abstract}

\date{Version 16.0 -- \today}

\maketitle

\section{Introduction}

Strongly correlated electronic systems with flat bands have shown several novel phenomena, including unconventional magnetism and superconductivity\cite{Hasan_2015,Cao2018insul,PhysRevLett.128.087002,PhysRevLett.128.096601}.
In particular, the experimental ``magic angle'' introduces the flat band into the graphene twisted lattice\cite{Cao2018super,PhysRevB.99.235417}, representing one of the most promising Dirac fermion systems\cite{GUPTA201544,Geim2007} which has received much attention\cite{Cao2021related,PhysRevLett.129.047601}.
Researchers now expect to reveal more physics and tune material properties through constructing band structures,
such as transport phase transition\cite{PhysRevB.99.155415,PhysRevLett.123.197702}, superconductor critical temperature\cite{Cao2018super,doi:10.1126/science.aav1910,PhysRevB.93.214505} and ferromagnetic (FM) order\cite{doi:10.1126/science.aaw3780,PhysRevLett.122.246402}.
However, it is challenging to obtain the appropriate angle through precise fine-tuning, which inspires the need for finding an alternative route to create the flat band in the lattice.
Although the Landau level under magnetic fields is a celebrated example among several methods\cite{PhysRevLett.60.2765,Du2009Landau,Bolotin2009Landau}, its effect of breaking time-reversal symmetry limits the emergence of certain correlated states\cite{Mao2020FB}.
Applied to materials with better electronic and mechanical properties appearing\cite{BRANDAO2023104503,Shu2022graphene}, experimental strain has been an efficient technique to enhance device performance through buckling, tearing or folding lattice layers\cite{PhysRevMaterials.6.L041001,doi:10.1146/annurev-matsci-082908-145312}. Periodic deformation is proven to be able to couple with electronic motion and masquerade as a valley-antisymmetric pseudomagnetic field\cite{PhysRevLett.128.176406}, therefore inducing pseudo-Landau levels\cite{Mao2020FB,PhysRevB.100.035448} characterized by the nonuniform distribution of carrier density in real space\cite{Mao2020FB,Kang2020}. It provides a simple and feasible way to accumulate and investigate the flat-band system\cite{PhysRevMaterials.6.L041001}.

Single-layer graphene (SLG) is one of the most promising materials due to exhibiting a range of interesting phenomena\cite{Geim2007,GUPTA201544}, among which transport properties attract considerable interest\cite{PhysRevLett.104.136803SLGSC,doi:10.1126/science.1144359}.
For the experimental evidence indicating the potential of this material in related fields\cite{DiBernardo2017,PhysRevB.80.094522,PhysRevB.90.245114}, a feasible method to reveal and investigate superconducting behavior in graphene monolayers requires urgent development.
Outside of doping carriers which has been proven to be reliable in some two-dimensional (2D) materials\cite{RevModPhys.78.17,Phillips2020},
recently, some experiments proposed the novel strain array in SLG, which reconstructed the energy bands into a series of essentially flat bands\cite{Mao2020FB,Jiang2017},
demonstrating an interesting technique for strain to induce symmetry breaking and then play a central role in attractive magnetism and superconductivity\cite{PhysRevLett.111.046604,PhysRevX.4.021042,PhysRevB.96.064505,Park2022,PhysRevB.83.220503}.
In fact, the influence of strain on spontaneous symmetry breaking has shown a close correlation with attractive phenomena\cite{doi:10.1126/science.1248292,PhysRevB.90.041413,PhysRevB.104.054518}.
In particular, this effect could avoid the limitation that the interaction in real material is not strong enough to induce a Mott insulator\cite{PhysRevLett.106.236805,PhysRevLett.120.116601},
which may provide fertile ground for hosting strongly correlated phenomena and help predict possible further development in experiments.

In this paper, we studied the Hubbard model on the SLG through the exact determinant quantum Monte Carlo (DQMC) method.
Based on experiments on pseudo-Landau levels induced by deformation\cite{Mao2020FB}, we adjusted the electron hopping to design a 2D strain lattice.
In this system, the nonuniform electron distribution in real space verifies flat bands in momentum space.
Choosing the proper chemical potential, we control the charge density in the constructed flat band range. Through the change in pairing susceptibility and magnetic susceptibility for different interactions and strain strengths,
we suggest that the strain enhances the superconductivity in flat band with the next-nearest neighbor $d+id$(in the following briefly denoted $N\text{-}d+id$) pairing symmetry dominating others, and the system exhibits significant ferromagnetism, which is attributed to symmetry breaking.
We also use the magnetic structure factor and density of states at the Fermi energy to further investigate the effect of strain on symmetry breaking at half-filling.
We found that the induced energy-gap antiferromagnetic (AFM) insulator appears at interaction $U/t=3$ close to the situation of real material and has similar characteristics to the Mott insulator under strong Coulomb repulsion.
Although the sign problem and error bar limit the range of charge density, our discussion on the influences of the flat band and the comparison with experiments remain valuable.

\section{Model and method}

\begin{figure}[b]
\includegraphics[scale=0.4]{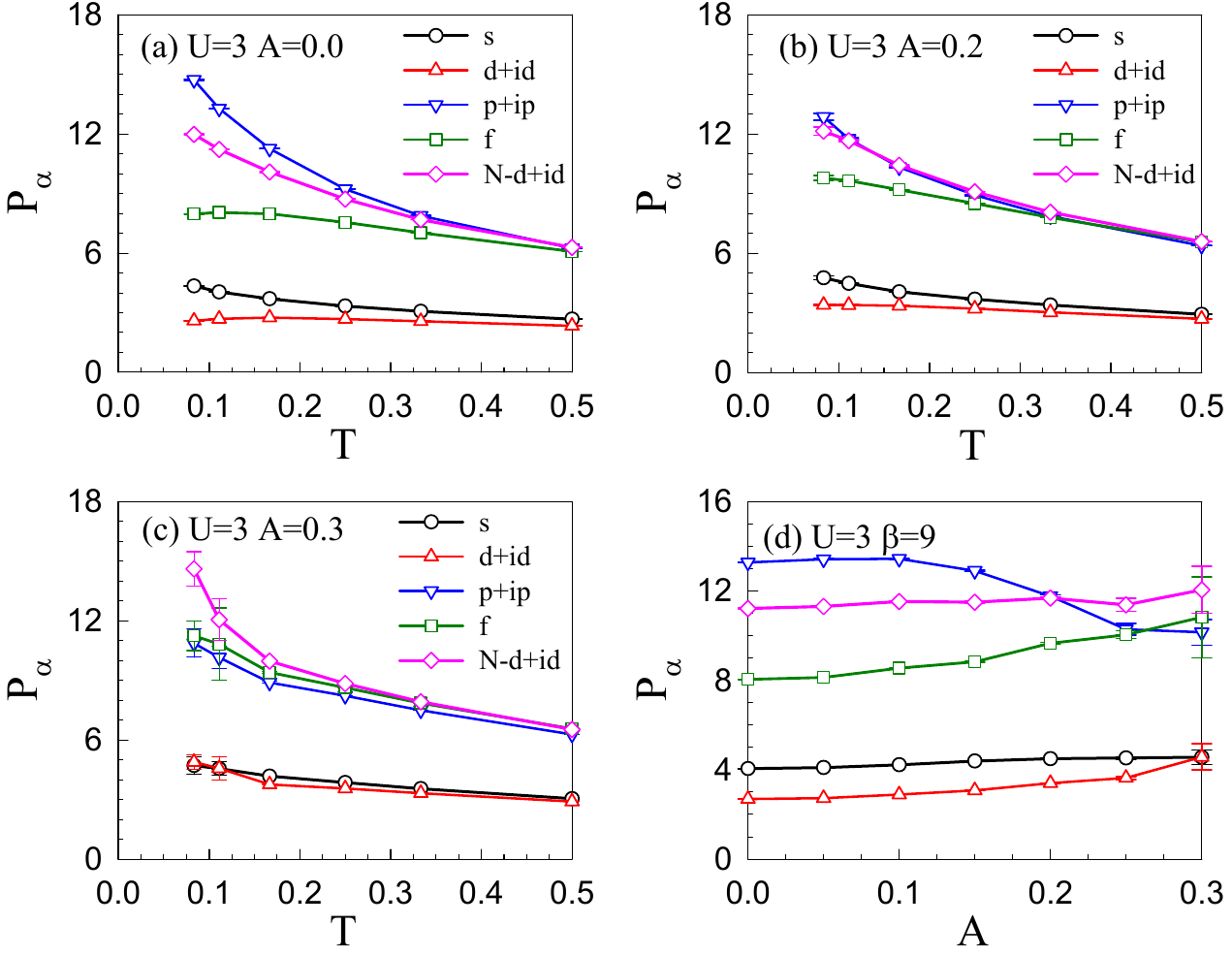}
\centering
\caption{\label{FigPT} The pairing susceptibility $P_{\alpha}$ computed as a function of temperature $T$ for strain strength $A$ equal to: (a) 0.0, (b) 0.2, (c) 0.3. As $T$ decreases, $P_{N\text{-}d+id}$ has a divergent tendency under the pretty strong $A$. (d) $P_{\alpha}$ as a function of $A$ at inverse temperature $\beta=1/T=9$. The dominant pairing changes at $A\sim0.2$ where there is an intersection. For $A=$0.2 and 0.3, the average changes in hopping parameters are respectively 0.239 and 0.367. Error bars under weak strain are pretty small.
}
\end{figure}

The Hamiltonian of the interacting Hubbard model on a honeycomb lattice in the presence of asymmetrical strain is defined as follows:
\begin{eqnarray}
\label{Hamiltonian}
\hat{H}&=&-\sum_{\langle{\bf ij}\rangle\sigma}t_{ij}(\hat{c}_{{\bf i}\sigma}^\dagger\hat{c}_{{\bf j}\sigma}
+\hat{c}_{{\bf j}\sigma}^\dagger\hat{c}_{{\bf i}\sigma})+U\sum_{\bf j}(\hat{n}_{{\bf j}\uparrow}-\frac{1}{2})(\hat{n}_{{\bf j}\downarrow}-\frac{1}{2}) \nonumber\\
&&-\sum_{{\bf j}\sigma}\mu\hat{n}_{{\bf j}\sigma}.
\end{eqnarray}
In Eq.~\eqref{Hamiltonian}, $\hat{c}_{{\bf i}\sigma}^\dagger(\hat{c}_{{\bf i}\sigma})$ is the spin-$\sigma$ electron creation (annihilation) operator at site $\bf i$, and $\hat{n}_{{\bf i}\sigma}=\hat{c}_{{\bf i}\sigma}^\dagger\hat{c}_{{\bf i}\sigma}$ is the occupation number operator. Here, $U > 0$ is the onsite Coulomb repulsive interaction, and $\mu$ is the chemical potential.
To build the strained graphene lattice, we defined the \textit{nearest-neighbor} hopping integral $t_{ij}$ as:
\begin{eqnarray}
\label{tij}
t_{ij}=t_{0}e^{-\beta_{0}\frac{\delta_{a}}{a_{0}}},
\end{eqnarray}
where $t_{0} = 1$ sets the energy scale, $a_{0}$ is the carbon$\text{-}$carbon distance, $\beta_{0}$ is approximately 3 in graphene\cite{PhysRevB.100.205411,PhysRevB.86.125402}, and $\delta_{a}$ is the change in interatomic distance under strain.
We include the deformations, which is on $z-$direction perpendicular to the lattice plane, referring to Ref. \cite{Mao2020FB} given by
\begin{eqnarray}
\label{dz}
z(\mathbf{r_{i}})=A_{0}[\cos(\mathbf{b_{1}}\cdot \mathbf{r_{i}})+\cos(\mathbf{b_{2}}\cdot \mathbf{r_{i}})+\cos(\mathbf{b_{3}}\cdot \mathbf{r_{i}})].
\end{eqnarray}
In Eq.~\eqref{dz}, $A_{0}$ is the amplitude of periodic strain, which is used to describe the strength of strain. $\mathbf{b_{1}}=\frac{2\pi}{a_{b}}(0,1)$, $\mathbf{b_{2}}=\frac{2\pi}{a_{b}}(-\frac{\sqrt3}{2},-\frac{1}{2})$, $\mathbf{b_{3}}=-(\mathbf{b_{1}}+\mathbf{b_{2}})$, and $a_{b}$ here is equal to $a_{0}$, which sets the strain period.
If two sites have different deformations on the $z-$direction and we define the difference as $dz$, then the new distance between these two sites becomes $a'=\sqrt{dz^{2}+a_{0}^2}$, therefore $\delta_{a}=\sqrt{dz^{2}+a_{0}^2}-a_{0}$ and the new hopping parameter $t'$ becomes $t_{0}e^{\sqrt{1+(dz/a_{0})^{2}}-1}$.
For the sake of representation, we define strain strength $A$ as $ A_{0}/(\sqrt{3}a_{0}/2)=A_{0}/a_{1}$, in which $a_{1}=\sqrt{3}a_{0}/2$ is shown in Fig.~\ref{FigPT}(f).
In Fig.~\ref{FigPT}(f), hopping parameters between blue and red sites become $t'$, while hopping parameters between blue and blue sites are still $t_{0}$.
Our numerical calculations are performed on a graphene ribbon with periodic boundary conditions by using the DQMC method, and the lattice size $N_{s}$ is $2\times L^{2}$, in which $L=8$ is the number of cells in each direction. The DQMC simulation is a powerful unbiased numerical tool to study strongly correlated systems, whose basic strategy is to express the partition function as a high-dimensional integral over a set of random auxiliary fields\cite{PhysRevLett.62.961,PhysRevB.69.104504,RevModPhys.84.1383,PhysRevLett.111.066804}.
The origin of the strain period is located at the center of the lattice, and the deformation reconstructs the band structure into a series of flat bands, as shown in Fig.~\ref{SMEk}. The charge density $n$ set to 0.4 is in the range of the flat band, which is the second closest band to the $E=0$ level.

\begin{figure}[t]
\includegraphics[scale=0.8]{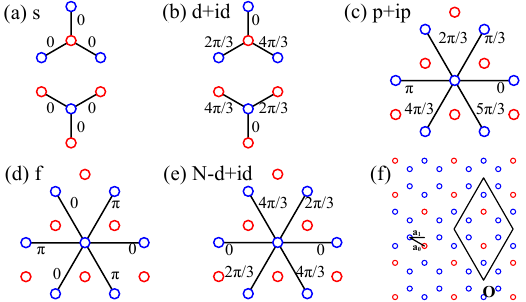}
\centering
\caption{{Phases and $\delta_{l}$ values of (a) $s$ (b) $d+id$ (c) $p+ip$ (d) $f$ and (e) $N\text{-}d+id$ pairing symmetry. (f) The form of the applied strain, and $\alpha$ or $\beta$ sites are labeled by color red or blue.}
}
\label{pair}
\end{figure}

To study magnetism, and to particularly characterize ferromagnetism, we compute the spin susceptibility as
\begin{eqnarray}
\label{Xq}
\chi(q)=\int_{0}^{\beta}d\tau\sum_{j,k} e^{i\mathbf{q}\cdot(\mathbf{r_{j}}-\mathbf{r_{k}})} \langle m_{j}(\tau)\cdot m_{k}(0)\rangle,
\end{eqnarray}
where $m_{j}(\tau)=e^{H\tau}m_{j}(0)e^{-H\tau}$ with $m_{j}=\hat{c}_{{\bf i}\uparrow}^\dagger\hat{c}_{{\bf i}\uparrow}-\hat{c}_{{\bf i}\downarrow}^\dagger\hat{c}_{{\bf i}\downarrow}$. We also define the pairing susceptibility to investigate the superconductivity in a strained graphene lattice:
\begin{eqnarray}
\label{Pa}
P_{\alpha}=\frac{1}{N_{s}}\sum_{\mathbf{i},\mathbf{j}}\int_{0}^{\beta}d\tau \langle \Delta_{\alpha}^\dagger(\mathbf{i},\tau)\Delta_{\alpha}(\mathbf{j},0)\rangle.
\end{eqnarray}
Here, $\alpha$ denotes the pairing symmetry, and the corresponding pairing order parameter $\Delta_{\alpha}^\dagger(i)$ is defined as
\begin{eqnarray}
\label{PaD}
\Delta_{\alpha}^\dagger(i)\equiv\sum_{l}f_{\alpha}^{\ast}(\delta_{l})(c_{i\uparrow}c_{i+\delta_{l}\downarrow}\pm c_{i\downarrow}c_{i+\delta_{l}\uparrow})^{\dag},
\end{eqnarray}
where $f_{\alpha}(\delta_{l})$ is the form factor of the pairing function, the vectors $\delta_{l}$ denote the bond connections, and ``$-$''(``+'') labels singlet (triplet) symmetry. The vectors $\delta_{l}$ in which $l=$1, 2, 3 (or $l=$1, 2, 3, 4, 5, 6) represent the nearest (or next-nearest) connections. More details are shown in Fig.~\ref{pair}.

We define $N(0)$, the density of states at the Fermi level, as
\begin{eqnarray}
\label{N0}
N(0)\simeq \beta \times G({\bf r} = 0,\tau = \beta/2)/\pi,
\end{eqnarray}
to characterize the energy gap at half filling. In Eq.~\eqref{N0}, $G$ is the imaginary-time-dependent Green's function. We finally introduce the magnetic structure factor as $S(q) = \sum_{j,k} e^{i\mathbf{q}\cdot(\mathbf{r_{j}}-\mathbf{r_{k}})} \left({S_{j}\cdot S_{k}}\right)$, where $S_{j}$ is the spin operator. The normalized AFM spin structure factor is defined as $S_{AFM}=S(K)//L^{2}$ to discuss the influence of the finite size effect, in which $K$ represents the $K\text{-}$point in the Brillouin zone of the graphene lattice.

\section{Results and Discussion}

Beginning with the pairing susceptibility $P_{\alpha}$ as a function of temperature $T$, we suggest that strain efficiently enhances the $N\text{-}d+id$ superconductivity.
Figure~\ref{FigPT} shows the behavior of $P_{\alpha}$ for different pairing symmetries at $U=3$ and $n=0.4$, in which the intensities of the $p+ip$, $f$ and $N\text{-}d+id$ pairing symmetries are much larger than those of the others and significantly increase with decreasing temperature. For the nonstrained system with dominant $p+ip$ pairing, introducing strain into the system will significantly suppress $P_{p+ip}$ while enhancing $P_{N\text{-}d+id}$. When the strength of strain $A$ exceeds 0.2, the $P_{N\text{-}d+id}$ curve will be lifted above the $P_{p+ip}$ curve and shows a divergent tendency as $T$ decreases. As shown in Fig.~\ref{FigPT}(d), the critical value is located at approximately $A=0.2$, where the two curves intersect.
We consider that this phenomenon is induced by the flat band near $n=0.4$, as shown in Fig.~\ref{SMEk} for the noninteracting condition. Weak dispersion and narrow band width lead interactive effects to dominate over those of kinetic energy, which leads us to propose this system as a platform for demonstrating strongly correlated phenomenon\cite{Balents2020}.

\begin{figure}[t]
\includegraphics[scale=0.6]{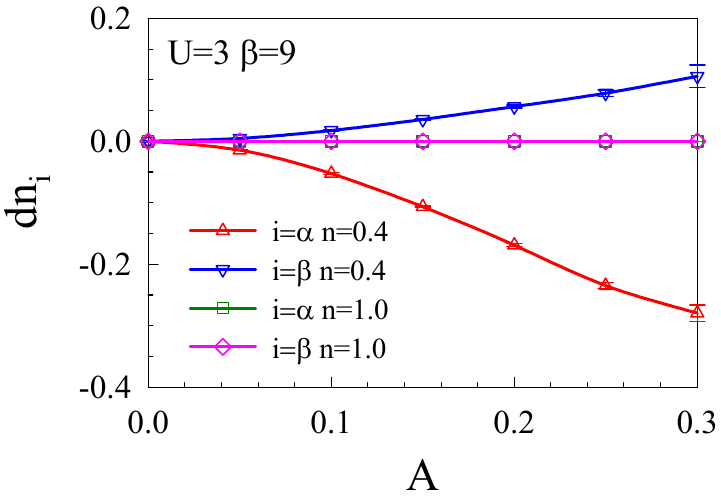}
\centering
\caption{\label{FigFB} $dn_{i}$, the difference value between electron density on site and average electron density, as the function of $A$. $i=\alpha$ or $\beta$ represents sites with different environment. $dn_{\alpha}$ and $dn_{\beta}$ gradually separate with increasing $A$ at $n=0.4$ and coincide at $n=1$.}
\end{figure}

We further confirm whether the flat band is stable and effective under the Coulomb interaction. In Fig.~\ref{FigFB} we use the nonuniform distribution of carrier density on lattice sites to illustrate the localization of electrons in real space and then the flat band in $k-$space.
Figure~\ref{pair}(f) shows the environment of origin point $O$ for periodic strain, the lattice deformation and the unit cell labeled by the black line. Under the strain shown in Eq.~\eqref{dz}, red sites denoted $\alpha$ are lifted in the direction perpendicular to the lattice plane, while blue sites denoted $\beta$ are lowered. We set $dn_{i}=\langle n_{i}\rangle-\langle n\rangle$, in which $i=\alpha$ or $\beta$, as the longitudinal axis to show the change in electron density on the $\alpha$ and $\beta$ sites with increasing $A$ under different doping levels.
At $n=0.4$, the nonuniform distribution of electrons is indicated by different changing tendencies of the $dn_{\alpha}$ and $dn_{\beta}$ curves. When $A$ increases to 0.3, $n_{\alpha}$ tends to $0.4-0.3=0.1$; therefore, electrons in the system are concentrated on $\beta$ sites. Compared with the half-filled condition in which overlapping curves of $dn_{\alpha}$ and $dn_{\beta}$ indicate a uniform electron distribution, we propose that if electrons are doped into the $n=0.4$ system until $n$ arrives at 1, it is the flat band near $E=0$ that would be filled (more details are in Fig.~\ref{SMFB}), and these added electrons are mainly localized on $\alpha$ sites. That is, the flat band in $k-$space would behave as electron localization in real space, and this result has also been shown in previous experimental and numerical studies\cite{Kang2020,Mao2020FB}.

\begin{figure}[t]
\includegraphics[scale=0.61]{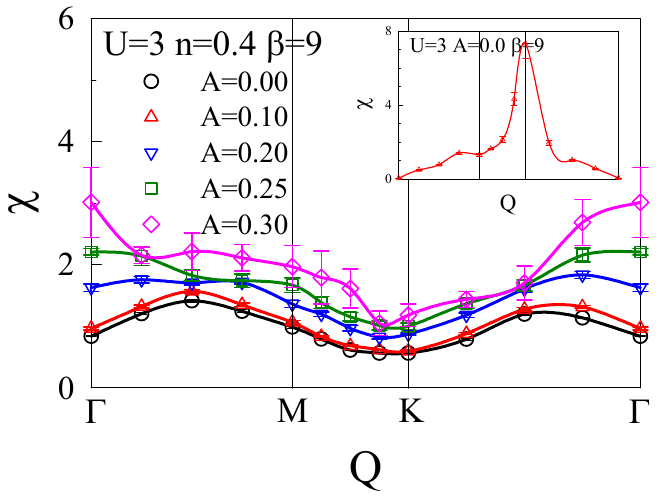}
\caption{\label{FigX}
Magnetic susceptibility $\chi$ in the Brillouin zone. As $A$ increases, the peak appearing at the $\Gamma$ point indicates FM. The insert shows the AFM at half-filling. For $A=$0.10, 0.20, 0.25 and 0.30, the average changes in hopping parameters are respectively 0.080, 0.239, 0.310 and 0.367.}
\end{figure}

In practice, a flat band with a high degeneracy of electronic energy levels is always accompanied by symmetry breaking and an FM state in a correlated system\cite{PhysRevB.85.085209,PhysRevLett.111.126804,doi:10.1021/acs.nanolett.1c01972,AMielke1991,PhysRevLett.69.1608}. Here, we use the $k-$space magnetic susceptibility $\chi$ to indicate that there is an actual tendency from antiferromagnetism to ferromagnetism.
We initially verify the AFM of the half-filled clean system in the inset of Fig.~\ref{FigX}, in which the $\chi(q)$ curve has a peak at the $K$ point. Doping holes until $n=0.4$, $\chi(q)$ at high symmetry points are not prominent. As $A$ grows to over 0.2, the changed susceptibility pattern has a significantly enhanced peak at the $\Gamma$ point, suggesting FM in the strained system.
Therefore, transitions of $P_{\alpha}$ and $\chi$ indicate the particularity of strength $A\sim0.2$, and strain building up the flat band is proven to induce novel correlated phenomena in the SLG system.
Notably, the critical strength $A=0.2$ yields the amplitude of periodic strain $A_{0}$ equal to $0.1\sqrt{3}a_{0}$, so the change in carbon-carbon distance $\delta_{a}/a_{0}$ is approximately 12.7$\%$. Compared to strain of 25.0$\%$ which can be achieved in graphene\cite{science1157996,Shu2022graphene}, we suggest our results are meaningful to studies on real materials.
We also show the finite-size analysis in Fig.~\ref{SMLeffect}.

As the appearing ferromagnetism suggests the possibility of symmetry breaking,
considering the half-filled case, here, we use the magnetic structure factor $S$ and density of states at the Fermi energy $N(0)$ to confirm the effect of strain on symmetry breaking, which may be related to transitions of superconductivity and magnetism. We perform an equal-time calculation to obtain a larger lattice size $L=12$, which is meaningful for obtaining accurate results, especially for the energy bands around the Fermi energy.

\begin{figure}[t]
\includegraphics[scale=0.4]{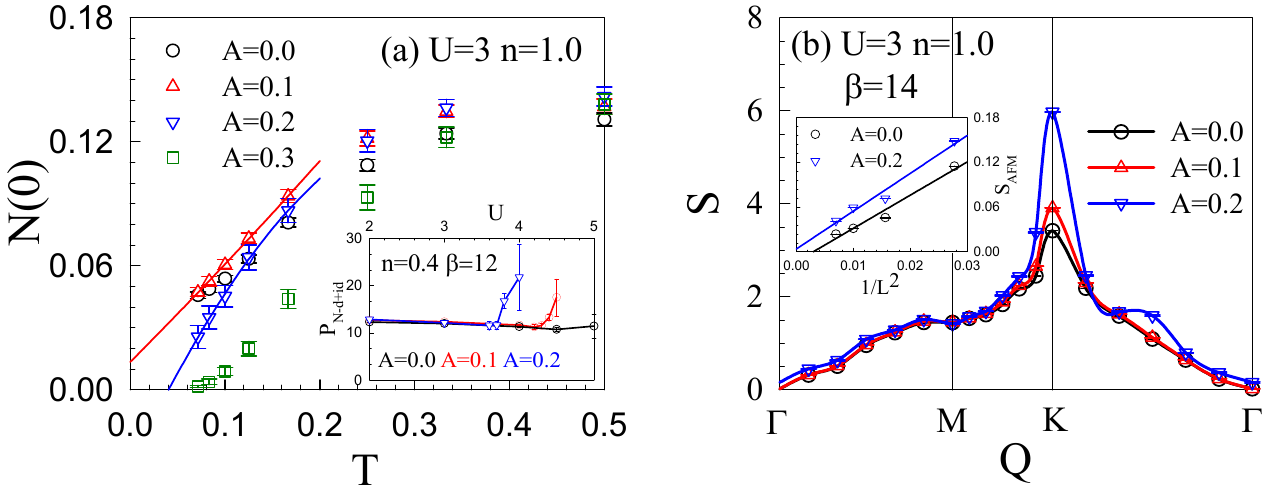}
\caption{\label{Figsym} (a) The density of states at the Fermi energy $N(0)$ as a function of temperature $T$. As $A$ increases, the $N(0)$ curve tends to be a negative value at $T\rightarrow0$, indicates an opened gap. The inset shows $P_{\alpha}$ as a function of $U$, which would be effectively lifted when the interaction exceeds the critical value $U_{c}(A)$. (b) Magnetic structure factor $S$ in the Brillouin zone. Increasing $A$ enhances the peak of cure at the $K$ point. The inset shows the normalized magnetic structure factor $S_{AFM}=S(K)//L^{2}$ as a function of lattice size $L$. The positive intercept of the curve represents the presence of AFM order.}
\end{figure}

Figure~\ref{Figsym} shows $N(0)$ as a function of $T$ under various $A$. The strain-free $N(0)$ curve has a positive intercept on the longitudinal axis, indicating that the density of states is finite at the Fermi energy under the interaction\cite{PhysRevLett.120.116601}. As the strain continues to increase, the $N(0)$ curve has a negative intercept under $A=0.2$ and decreases to zero under $A=0.3$ as $\beta$ tends to 14, which means that a band gap is opened at $E=0$.
Furthermore, we plot the magnetic structure factor $S$ of the $k-$space in Fig.~\ref{Figsym}(b), and the increasing $A$ significantly enhances the peak at the $K$ point, which represents the antiferromagnetic behavior that occurs in the graphene.
The inset of Fig.~\ref{Figsym}(b) shows the result of scaling to discuss the finite size effect, in which the normalized magnetic structure factor $S_{AFM}$ is a function of lattice size $L$.
Without strain, $S_{AFM}$ tends to 0 when $1/L\rightarrow0$ (namely, $L\rightarrow\infty$), which suggests that the structure
factor is not extensive, thereby resulting in only short-range ordering. When $A$ increases to 0.2, the $S_{AFM}$ curve is predicted to be positive at $1/L\rightarrow0$, corresponding to the AFM phase\cite{PhysRevB.105.205121}.

Therefore, the strain provides an AFM insulator with an energy gap at half-filling (a detailed transport result is shown in Fig.~\ref{SMsym}), although the limitation of $U$ equal to 9.3 eV/2.8 eV$\sim3.3$ from experiments always makes it difficult to obtain a Mott insulator\cite{PhysRevLett.106.236805,PhysRevLett.120.116601}. Doping holes until reaching a concentration of $n=0.4$, we show in the inset of Fig.~\ref{Figsym}(a) about $P_{N\text{-}d+id}$ as the function of $U$ under various $A$. Compared with the $A=0$ case where the curve would be slightly lifted at $U>4$, in the presence of strain, there is an obvious tendency for $P_{N\text{-}d+id}$ to increase rapidly at the critical point $U_{c}$, and $U_{c}$ decreases with increasing $A$. Therefore, it can be indicated that strain enhances the emergence of superconductivity in correlation and is accompanied by symmetry breaking.

\begin{figure}[t]
\includegraphics[scale=0.4]{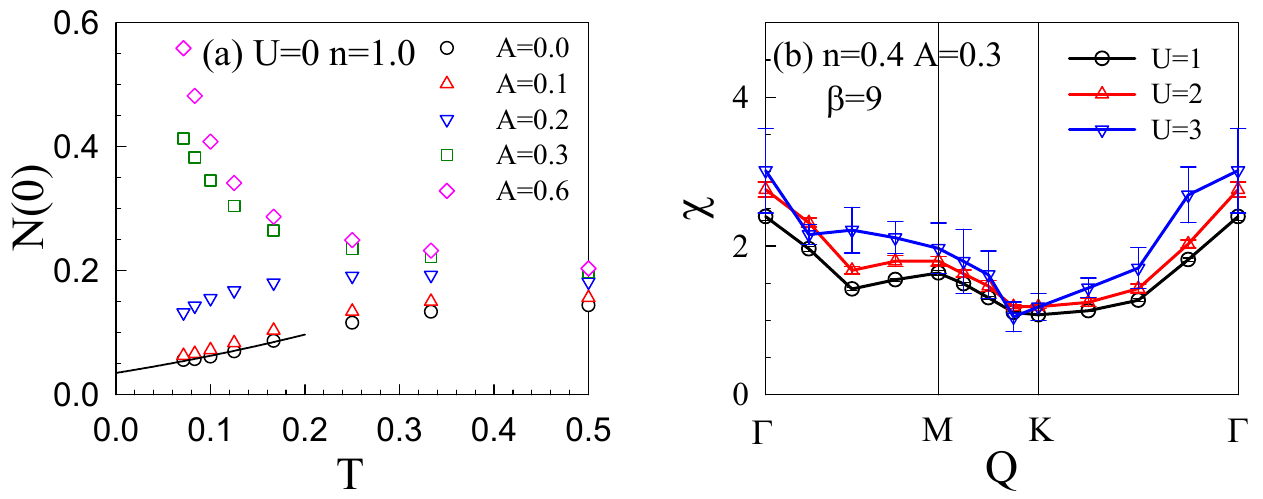}
\caption{\label{FigU} (a) $N(0)$ as a function of $T$. Even under a sufficiently strong $A$, the intercept of the curve is still positive, and the gap is absent.  For $A=$0.1, 0.2, 0.3 and 0.6, the average changes in hopping parameters are respectively 0.080, 0.239, 0.367 and 0.490. (b) $\chi$ in the Brillouin zone. Ferromagnetism would be enhanced by interaction in the strained system.}
\end{figure}

However, due to the presence of interactions and their interplay with strain, further work is needed to determine the role of strain in this effect.
We initially show the half-filled case without interaction. Figure~\ref{FigU}(a) shows $N(0)$ as a function of $A$, and even under a sufficiently strong $A=0.6$, the intercept of the $N(0)$ curve is still positive, so the gap is absent. We discuss in the Appendixes the strain strengths. Additionally, strain has almost no influence on magnetism, as shown in Fig.~\ref{SMsym}, and the AFM transition does not exist. Therefore, the influence of strain on symmetry only takes effect when $U\neq0$.
Then, we return to the doping case. As shown in Fig.~\ref{FigU}(b), increasing $U$ in the strained system will result in an increase in $\chi(q)$ at the $\Gamma$ point when $n=0.4$. That is, the FM appearing in the flat band range has a positive response to interaction, and $U$ and $A$ cooperate to induce symmetry breaking.

\section{Summary}

Using DQMC simulations, we studied transitions of superconductivity and magnetism accompanied by symmetry breaking in the Hubbard model on a strained graphene monolayer. At half-filling, strain accumulates an antiferromagnetic insulating state with an energy gap, and its effect of constructing flat bands induces the $N\text{-}d+id$ superconductivity with ferromagnetism when doping holes drives the electronic density into the flat-band range. The effect of strain on energy bands is indicated by the localization of electrons in real space, and the appearance of symmetry breaking is indicated by magnetic transition.

Finite interaction in SLG limits the possibility of introducing superconductors through doping a Mott insulator. Based on this observation, we propose strain as a method available for reference, which would induce a rapid enhancement of next-nearest neighbor $d+id$ pairing susceptibility when the interaction exceeds the critical value $U_{c}(A)$. Although the range of charge density is limited as shown in Fig.~\ref{SMsign}, discussions about the strain-induced flat band and its effect on transport and magnetism are still meaningful.

Last but not least, here we mainly focus on the out-of-plane deformation, which could be achieved when buckling of the membrane reduces elastic energy and results in intriguing periodic strain patterns\cite{science.1220335}, or it could be produced experimentally on suitable substrates\cite{Mao2020FB}.	We would like to mention that tension or defects, inducing in-plane deformation, provides another effective method to cause the change of hopping constants, while it is much different from the out-of-plane deformation. Some of us authors have investigated the metal-insulator transition induced by in-plane deformation, considered another realistic situation for in-plane part to be dominant\cite{LFZhangStrain}.

\noindent
\underbar{\bf ACKNOWLEDGMENTS:}
This work was supported by Beijing Natural Science
Foundation (No. 1242022) and NSFC (Nos. 12474218 and 12088101). The numerical simulations were performed at the HSCC of Beijing Normal University and on Tianhe-2JK in the Beijing Computational Science Research Center.



\appendix

\section{APPENDIX A: THE FLAT BAND RANGE}
\label{app:FB}

\begin{figure}
\includegraphics[scale=0.35]{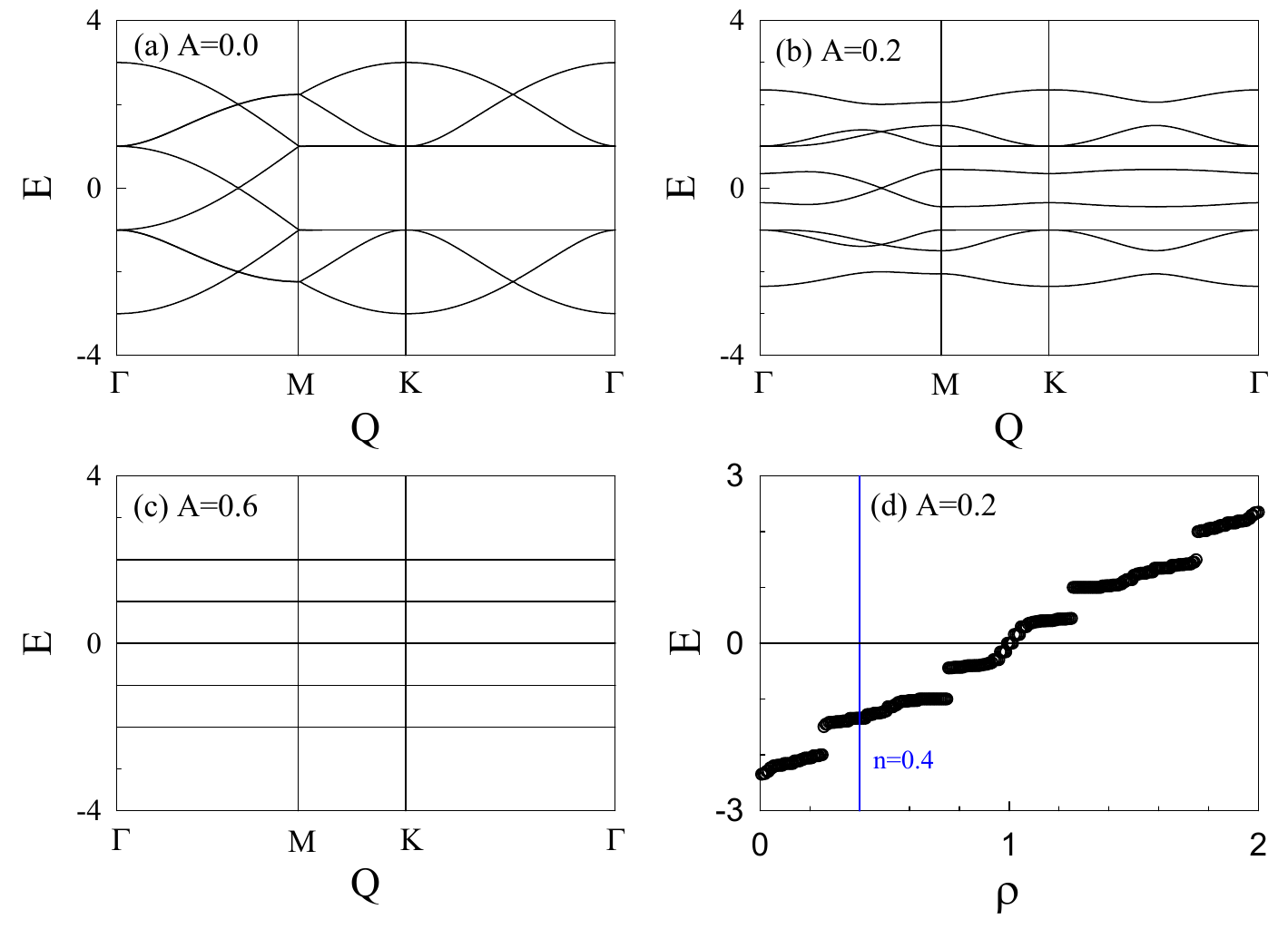}
\centering
\caption{\label{SMEk} Energy bands $E(k)$ at (a) $A=0.0$, (b) $A=0.2$, (c) $A=0.6$. As strain is gradually enhanced, energy bands are reconstructed into a series of flat bands. (d) Eigenvalues of the Hamiltonian of the $L=12$-strained lattice. The band gap is absent at the Fermi energy. The results are calculated in the noninteracting case.}
\end{figure}

\begin{figure}[b]
\includegraphics[scale=0.4]{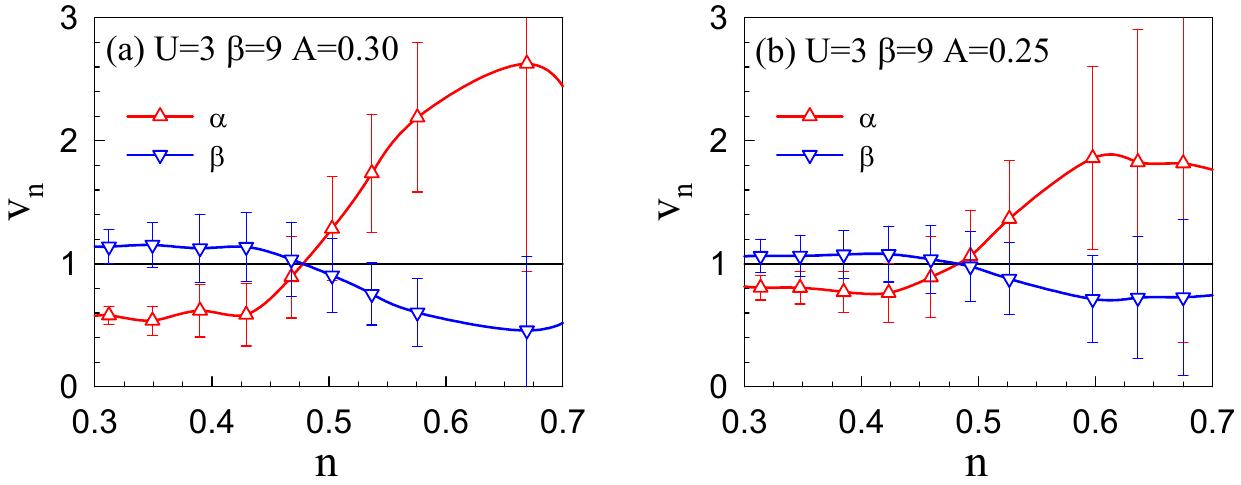}
\centering
\caption{\label{SMFB} The changing rate of density $v_{n}$ as a function of $n$ at (a) $A=0.30$ and (b) $A=0.25$. $v_{n}=1$ means that $n_{\alpha}$ or $n_{\beta}$ has the same change rate as the average density $n$. The crossing point of the $n_{\alpha}$, $n_{\beta}$ and $v_{n}=1$ curves separates the ranges of different flat bands.}
\end{figure}

We set the cell as shown in Fig.~\ref{pair}, which contains eight sites of different environments.
Through the second quantization, we obtain the energy bands, as shown in Figs.~\ref{SMEk}(a)-\ref{SMEk}(c), and the eigenvalues of the Hamiltonian matrix of the $L=12$ strained lattice are shown in Fig.~\ref{SMEk}(d). The increasing strength of strain accumulates a series of flat bands, and $A=0.6$ is large enough to reconstruct the energy bands. For flat bands around the Fermi energy, we choose $n=0.4$ to avoid the sign problem shown in Fig.~\ref{SMsign}.

In Fig.~\ref{FigFB} we use the localization of electrons to indicate the flat band in $k\_$space, especially the band near the Fermi energy. Compared to the case without interaction, here, we calculated the changing rate of density on sites $\alpha$ and $\beta$ as $v_{n}\sim \Delta n_{i}/\Delta n$ in Fig.~\ref{SMFB}. As the crossing of curves appears at $n\sim0.48$, the separation point of two flat bands is moved by interaction. When $n>0.48$, decreasing the chemical potential mainly decreases the charge density at $\alpha$ sites, which also indicates the localization of electrons in real space.

\section{APPENDIX B: THE FINITE SIZE EFFECT}
\label{app:FZ}

\begin{figure}[t]
\centerline {\includegraphics*[scale=0.36]{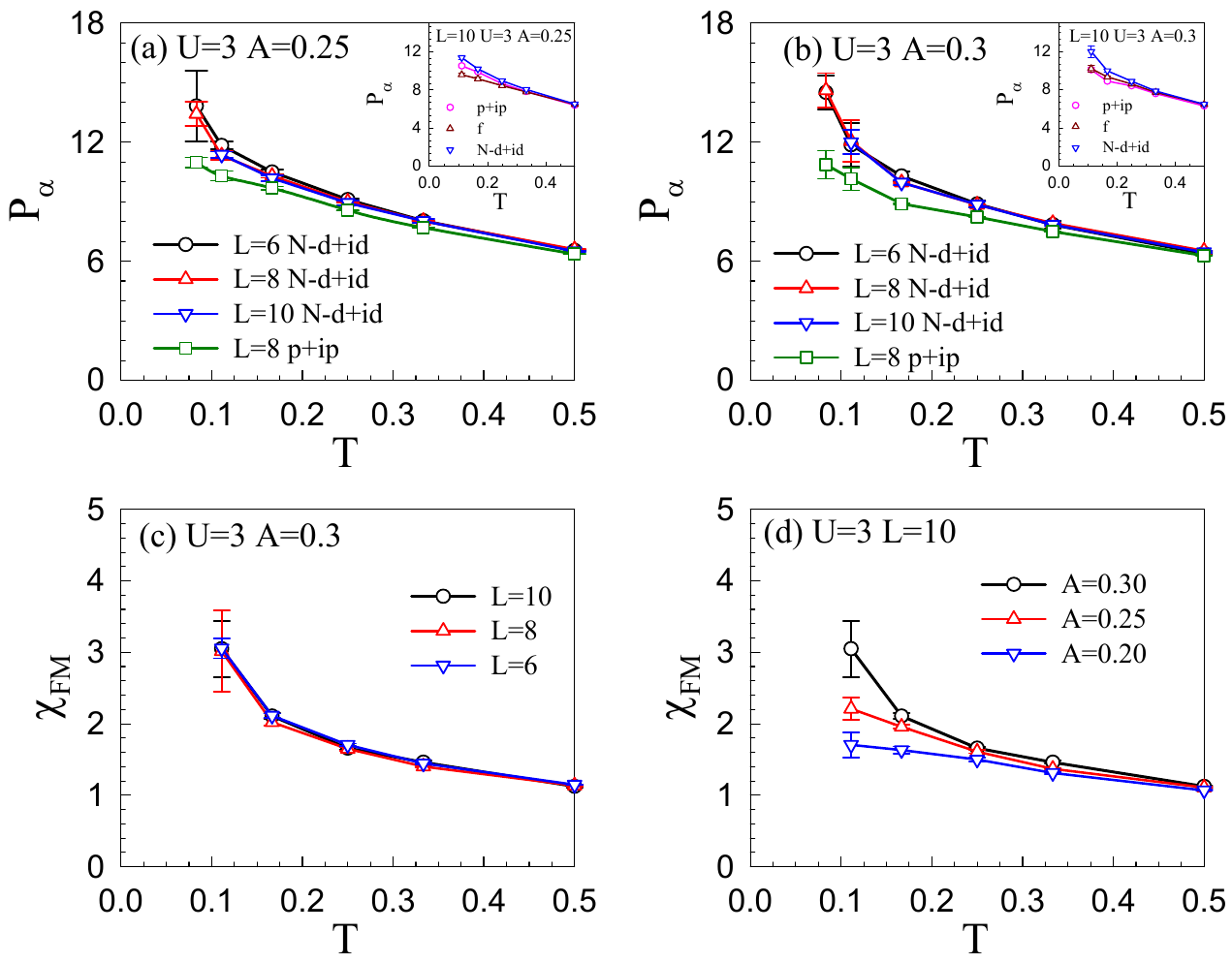}}
\caption{The pairing susceptibility $P_{\alpha}$ as a function of temperature $T$ with various lattice size $L$ at (a) $A=0.25$, (b) $A=0.30$. The main comparison is between $N-d+id$ wave and $p+ip$ wave. The insert shows $N-d+id$, $f$ and $p+ip$ pairing symmetries at $L=10$.
Ferromagnetic susceptibility $\chi_{FM}$ as a function of $T$ (c) with various $L$ at $A=0.3$ (d) with various $A$ at $L=10$. Calculations are based on $n=0.4$ system.}
\label{SMLeffect}
\end{figure}

To understand the influence of the system's finiteness on the physical results we have presented in the main text, here we report in Fig.~\ref{SMLeffect} the unequal-time pairing susceptibility and ferromagnetic susceptibility as the function of temperature $T$ for different lattice sizes.
We show in Figs.~\ref{SMLeffect}(a) and \ref{SMLeffect}(b) the $P_{N\text{-}d+id}$ on $L=6$, $L=8$ and $L=10$ lattices, in comparison with $P_{p+ip}$ on $L=8$ lattice. Curves with different lattice sizes basically coincide, verifying negligible finite-size effects, and $N\text{-}d+id$ pairing is dominant under each $L$. We also show pairing symmetries changed by strain at $L=10$ in inserts, and the phenomenon is similar to the one on $L=8$ lattice.

Then we check the magnetism. In Fig.~\ref{SMLeffect}(c), the ferromagnetic susceptibility $\chi_{FM}$ is the function of $T$, in which $\chi_{FM}=\chi(\Gamma)$ representing magnetic susceptibility at the $\Gamma\_$point in the Brillouin zone. With different lattice size $L$, $\chi_{FM}(T)$ curves have similar divergent tendency and very little difference, indicating the conclusion about FM under strong strain is pretty reliable. We also show $\chi_{FM}$ at $L=10$ with various $A$ in Fig.~\ref{SMLeffect}(d), which also indicates the enhancement of ferromagnetism by strain.

\begin{figure}[t]
\includegraphics[scale=0.4]{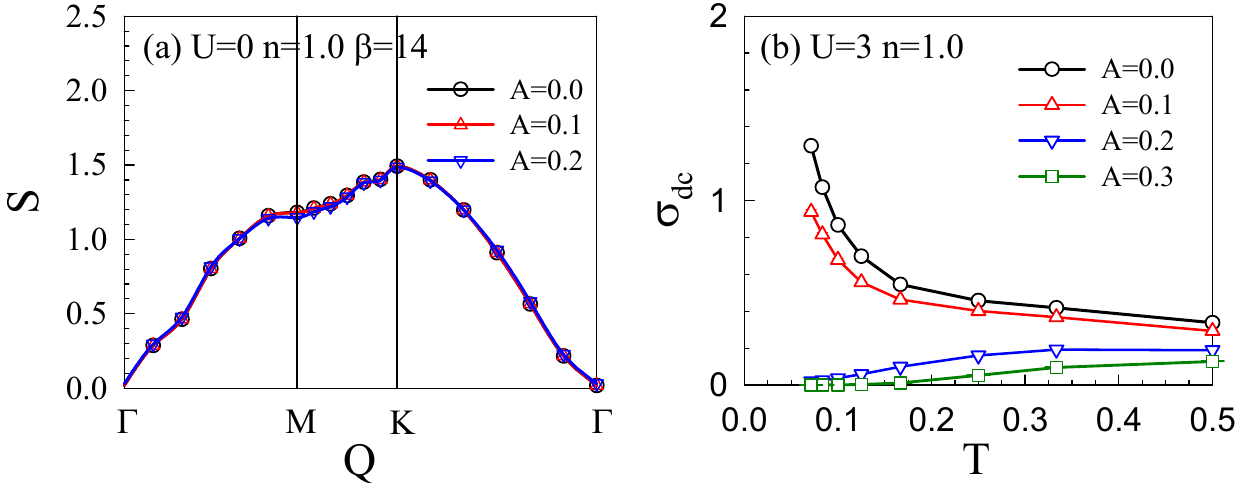}
\centering
\caption{\label{SMsym} (a) Magnetic structure factor $S$ in the Brillouin zone at $U=0$. The strain hardly changes the value of $S$. (b) Conductivity $\sigma_{dc}$ as a function of $T$. At low temperatures, $\sigma_{dc}(T)$ has different behaviors with $A$ greater/less than the critical strength.}
\end{figure}

\begin{figure}[b]
	\includegraphics[scale=0.4]{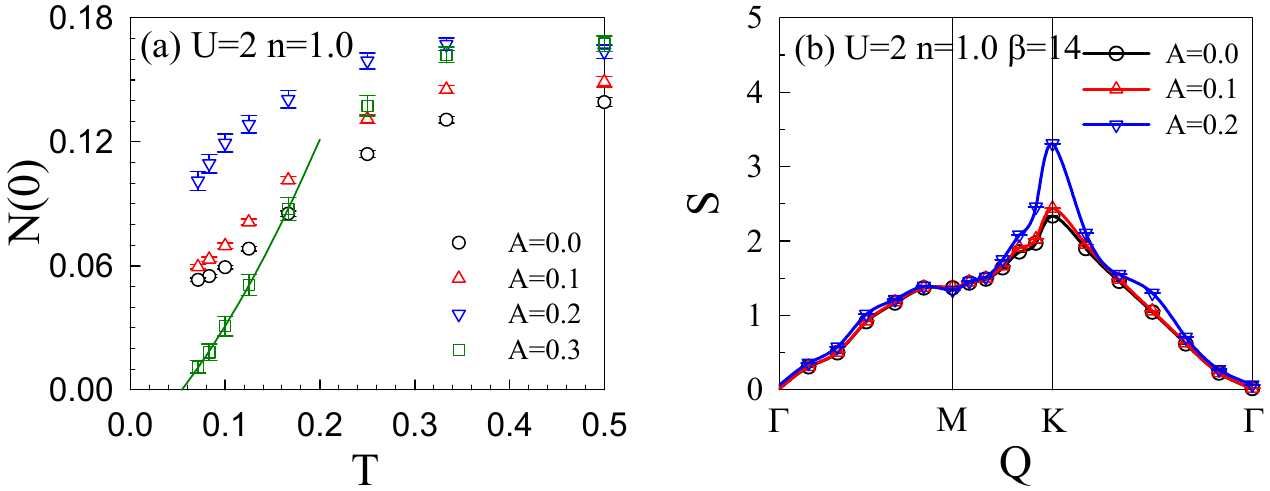}
	\centering
	\caption{\label{SMU2} The density of states at the Fermi energy $N(0)$ as a function of temperature $T$ at $U=2$. The critical value $A=0.3$, where the $N(0)$ curve tends to be a negative value when $T\rightarrow0$, is larger than the one at $U=3$. (b) Magnetic structure factor $S$ in the Brillouin zone at $U=2$.}
\end{figure}

\section{APPENDIX C: THE INTERACTION EFFECT AT HALF FILLING}
\label{app:n1Ueffect}

\setcounter{equation}{0}
\renewcommand{\theequation}{C\arabic{equation}}

To characterize the transport properties, we compute the $T$-dependent DC conductivity $\sigma_{\rm dc}$ via a proxy of the momentum $\bf q$ and imaginary time $\tau$-dependent current-current correlation function:
\begin{eqnarray}
\label{conduc}
\sigma_{\rm dc}(T)=\frac{\beta^{2}}{\pi}\Lambda_{xx}\left({\bf q}=0,\tau=\frac{\beta}{2}\right).
\end{eqnarray}
Here, $\Lambda_{xx}({\bf q},\tau)$ = $\langle \widehat{j_{x}}({\bf q},\tau)\widehat{j_{x}}(-{\bf q},0)\rangle$, and $\widehat{j_{x}}({\bf q},\tau)$ is the current operator. This form, which avoids the analytic continuation of the QMC data, has been shown to provide satisfactory results~\cite{PhysRevB.59.4364,doi:10.1073/pnas.1620651114,Mondaini2012}.

As shown in Fig.~\ref{SMsym}(b), we use the low-temperature behavior of $\sigma_{dc}$ to distinguish different transport properties. The half-filled system at $U=3.0$ and $A=0.0$ changes from metal ($\sigma_{dc}$ diverges with decreasing $T$) to insulator ($\sigma_{dc}$ decreases to 0 with decreasing $T$) as $A$ reaches 0.2. Therefore, considering the results of Fig.~\ref{Figsym}, the strain would induce an AFM insulating phase with an energy gap at $U=3$.
However, this effect is not always effective. As shown in Fig.~\ref{SMsym}(a), $A$ has little effect on magnetism at $U=0$.

\begin{figure}[t]
	\centerline {\includegraphics*[scale=0.45]{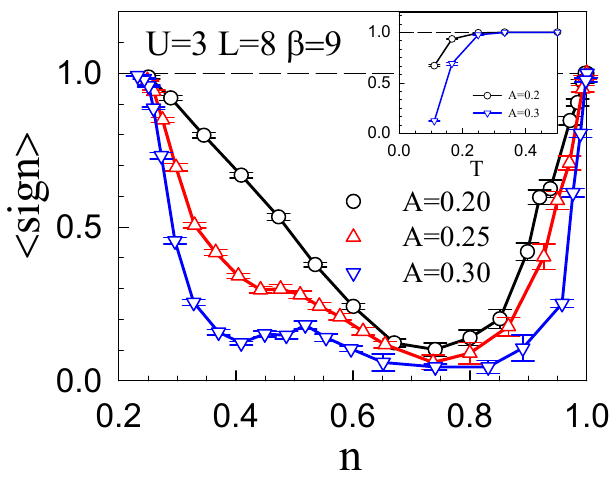}}
	\caption{\label{SMsign} The average fermion sign $\langle sign\rangle$ is shown as a function of charge density $n$ for various strain strength $A$ with interaction $U = 3$, lattice size $L = 8$, and inverse temperature $\beta = 9$. The inset shows $\langle sign\rangle$ as the function of $T$.}
\end{figure}

We propose that the strain and interaction would ``cooperate'' for symmetry breaking. As shown in Fig.~\ref{SMU2}, the critical strain of inducing the energy gap will increase when $U$ decreases from 3 to 2. Additionally, at half filling, it is more difficult for strain to enhance the AFM property.

\section{APPENDIX D: THE SIGN PROBLEM}
\label{app:sign}

\setcounter{equation}{0}
\renewcommand{\theequation}{D\arabic{equation}}

To characterize the sign problem, here we calculate the average fermion sign $\langle sign\rangle$ by computing the ratio of the integral of the
product of up and down spin determinants, to the integral of the absolute value of the product~\cite{PhysRevB.92.045110}:
\begin{eqnarray}
\label{sign}
\langle sign\rangle=\frac{\sum_{\chi}detM_{\uparrow}(\chi)detM_{\downarrow}(\chi)}{|\sum_{\chi}detM_{\uparrow}(\chi)detM_{\downarrow}(\chi)|}.
\end{eqnarray}
Here, $M_{\sigma}(\chi)$ represents each spin specie matrix. As shown in Fig.~\ref{SMsign}, when adjusting $\mu$ to decrease $n$ from 1, $\langle sign\rangle$ will decrease to around 0 rapidly, and float around 0 until $n \sim 0.65$. Then, as $n$ decreases to 0.2, $\langle sign\rangle$ gradually recovers to 1. The smaller the $\langle sign\rangle$, the more serious the sign problem, therefore when investigating the property of flat band, we should choose proper charge density to avoid the sign problem being too serious.
In the inset, it can be seen that $\langle sign\rangle$ is pretty large until $T$ decreases to 0.15, therefore the conclusion about dominant $N\text{-}d+id$ at $U=3$, $n=0.4$ and $A=0.3$ is reliable at high temperature. When $\langle sign\rangle$ is pretty small, we then improve the simulation parameters like warmup and measuring sweeps to ensure the accuracy.
To obtain the same quality of data as
$\langle sign\rangle \simeq$  1.0, indeed, one can estimate that the runs need
to be stretched by a factor on the order of ${\langle sign\rangle}^2$.

\bibliography{bib_ref}

\end{document}